\documentclass[12pt]{iopart}
\usepackage{iopams}



\long\def\comment#1{}

\def\W2{{\cal W}}

\def\be{\begin{equation}}
\def\ee{\end{equation}}
\def\bea{\begin{eqnarray}}
\def\eea{\end{eqnarray}}

\def\cmm2{{\,\rm cm^{-2}}}
\def\cm2{{\,{\rm cm}^2}}
\def\cmm3{{\,{\rm cm}^{-3}}}
\def\gcmm3{{\,{\rm g\,cm^{-3}}}}

\def\fun#1#2{\lower3.6pt\vbox{\baselineskip0pt\lineskip.9pt
  \ialign{$\mathsurround=0pt#1\hfil##\hfil$\crcr#2\crcr\sim\crcr}}}

\begin{document}
\title{Constraints from cosmic rays on non-systematic Lorentz violation}
\author{Sayandeb Basu\footnote{email: sbasu@physics.ucdavis.edu} and David Mattingly\footnote{email: mattingly@physics.ucdavis.edu}}
\address{Department of Physics, University of California, Davis, CA
95616, USA}

\begin{abstract}
In this article we analyze the radiation loss from a high energy
cosmic ray proton propagating in a spacetime with non-systematic
Lorentz violation.  From an effective field theory perspective we
illuminate flaws in previous attempts that use threshold
approaches to analyze this problem. We argue that in general such
approaches are of rather limited use when dealing with
non-systematic Lorentz violating scenarios.  The main issues we
raise are a) the limited applicability of threshold energy
conservation rules when translation invariance is broken and b)
the large amounts of proton particle production due to the time
dependence of the fluctuations. Ignoring particle production, we
derive a constraint on the magnitude of velocity fluctuation
$|v_f|<10^{-6.5}$, much weaker than has been previously argued.
However, we show that in fact particle production makes any such
constraint completely unreliable.

\end{abstract}

\section{Introduction}
Testing quantum gravity \emph{prima facie} seems to be an
impossible task, since the natural scale of the theory, the Planck
energy of $10^{19}$ GeV, is sixteen orders of magnitude in energy
beyond the reach of earth-based accelerators.  If we cannot
somehow gain knowledge about Planck scale physics, however, then a
complete theory of quantum gravity will most likely remain out of
reach. Fortunately, even if it is not possible to directly probe
Planck-scale physics, quantum gravity may leave tiny residual
effects at lower energies that are observationally testable.  A
promising avenue in this vein, and certainly the one that has
received the most attention, is the possibility that Lorentz
invariance is violated at the Planck scale.

If Lorentz symmetry is broken by quantum gravity, then it is only
an approximate symmetry and there should be small Lorentz
violating effects in low energy physics. There are two main
approaches to modelling low energy Lorentz violating effects.  The
first is the standard model extension \cite{Colladay:1998fq}. In
this approach one follows the tenets of effective field theory and
adds all possible Lorentz violating, renormalizable operators to
the standard model. This definite (and large) set of operators
yields a number of different effects, such as sidereal signals in
clock comparison experiments c.f. \cite{Bear:2000cd} or modified
neutrino oscillations \cite{Coleman:1997xq}, which can be used to
constrain the Lorentz violating coefficients in the standard model
extension. For a description of the current bounds on the standard
model extension see \cite{SecondCPT} and references therein.

An alternative approach is that of modified dispersion for
elementary particles, since one would expect Lorentz violating
corrections to the usual particle dispersion
$E^2=p^2+m^2$.\footnote{An effective field theory of course yields
particle dispersion relations. However, often a dispersion
modification alone is studied without explicitly working in a
particular field theory.} Usually in the literature rotation
invariance is assumed to hold in some frame, for the following
reason. If the rotation subgroup of the Lorentz group is broken
but the boost subgroup is preserved, then rotation breaking would
occur at all energies. Since rotation symmetry is a very good
symmetry at low energies, Planck scale rotation breaking is either
tiny or accompanied by broken boost invariance. The same is not
true of a violation of boost symmetry. We have explored only a
tiny fraction of the boost subgroup, which is after all unbounded.
Hence it is possible that boost Lorentz violation could be small
at low energies but become significant as the energies approach
the Planck scale. From a phenomenological standpoint is is
therefore logical to look for broken boost invariance first while
preserving rotation invariance.

Under the assumption of rotation invariance a dispersion
modification would typically look like
\begin{equation} \label{eq:disp}
E^{2}=p^2+m_A^2+\eta_A \frac{p^n}{M_{Pl}^{n-2}}
\end{equation}
where $\eta_A$ is usually taken to be an O(1) coefficient that can
depend on particle species $A$.  $\eta_A$ can also depend on
particle properties such as helicity, however we shall ignore such
possibilities as they are not relevant for our analysis. $n$ is an
integer determined by the underlying assumptions about quantum
gravity. Note that if $n>2$ then these dispersion modifications
would come from non-renormalizable terms in a particle Lagrangian.
Modified dispersion affects the kinematics of reactions, leading
to anomalous thresholds---reactions usually forbidden can be
allowed, and threshold energies for already allowed reactions can
be raised or lowered \cite{Jacobson:2002hd, Coleman:1997xq}.  The
usefulness of modified dispersion is that the energy at which the
reaction kinematics change is often far below the Planck energy.

To see this, consider the effect of Lorentz violation with $n=4$
on the existence of ultra-high energy cosmic rays (UHECR), as done
in~\cite{Gagnon:2004xh, Jacobson:2002hd, Coleman:1997xq}.  We
assume for this analysis that the UHECR are actually protons. If
$\eta_p>0$ then protons above a threshold energy of approximately
$E_{th}\approx(m_p^2 M_{Pl}^2/\eta_p)^{1/4}$ are superluminal and
lose energy rapidly via Cerenkov emission. In such a scenario
there would be a sharp break in the UHECR spectrum at $E_{th}$ as
there could be no long lived protons above $E_{th}$. For an
$\eta_p$ of O(1), $E_{th} \approx 10^{18}$ eV. Since we observe a
UHECR spectrum out to $10^{20}$ eV, $\eta_p$ is constrained to be
much less than O(1). So we see that we can constrain Lorentz
violating terms with current astrophysical measurements even if
those terms are Planck suppressed, as expected if they arise from
quantum gravity.

A modification of the form in eq.\ (1) is an example of systematic
Lorentz violation, which is constant over the lifetime of a
particle. However, some ideas about quantum gravity
\cite{Dowker:2003hb, Shiokawa:2000em, Ng:1993jb} give rise to
stochastic fluctuations in the world-line of a particle over its
lifetime.  The question then arises, what happens if we combine
the two ideas of Lorentz violation and non-systematic effects?  In
this paper we are concerned with the phenomenology of such
non-systematic Lorentz violation, especially with the
non-systematic version of the Cerenkov effect.\footnote{For a
discussion of the systematic Cerenkov effect, see for
example~\cite{Lehnert:2004be}.} There has been some previous work
on non-systematic scenarios and their possible observational
signatures. Amelino-Camelia \cite{Amelino-Camelia:2003zf} has
looked at possible imprints of metric stochasticity on gravity
wave interferometry. The basic conjecture is that fluctuations of
geometry induces a new noise-strain that is within the range of
detectability of space-based observatories such as LISA. However,
for this to happen it must be assumed in the model that
fluctuations add up coherently along the entire length of the
interferometer arms,\footnote{We thank Steve Carlip for pointing
this out.} which is a very fine tuned scenario that seems
unlikely.

Aloisio et. al \cite{Aloisio:2002ed,Aloisio:2004kh} and others
 \cite{Amelino-Camelia:2002ws, Jankiewicz:2003sm} have considered how non-systematic dispersion fluctuations affect
the propagation of ultra-high energy cosmic rays (UHECR) and
analyzed shifts in the thresholds for processes such as the GZK
reaction, $p+\gamma_{\textrm{\tiny{CMB}}}\rightarrow p+\pi^{0}$.
We shall focus on the conclusions of Aloisio et. al., although the
general arguments are applicable to any non-systematic threshold
analysis. In \cite{Aloisio:2002ed,Aloisio:2004kh} the authors use
a non-systematic dispersion relation similar to (\ref{eq:disp})
where the coefficient $\eta_A$ varies randomly as the particle
propagates. It is further assumed that the timescale of the
variation is set by the de Broglie frequency $1/E$ of the UHECR.
With these assumptions they then derive that the GZK-threshold
shifts to sub-GZK regimes and conclude that observations below the
GZK energy scale are enough to constrain scenarios of fluctuating
spacetimes. Furthermore they find a stability problem with UHECR,
namely that all charged particles are unstable to the vacuum
Cerenkov emission of photons. This conclusion, which is a primary
focus of this paper, seems encouraging at first sight, since
sub-GZK cosmic rays follow a simple power law spectrum and are
obviously stable over cosmological times. Fluctuating models as
considered by Aloisio et. al. would therefore seem to be ruled
out.

The analysis of Aloisio et. al. only considers the effects of
fluctuating dispersion relations on energy conservation. There is
a problem with this sort of analysis however - the length scale of
dispersion fluctuations may be shorter than the de Broglie
wavelength of other particles in the reaction (as pointed out in
\cite{Aloisio:2004up}). There are many fluctuations in the
interaction region and hence no single energy-momentum conserving
$\delta$-function.

Furthermore, fluctuations require some time-dependent background
field (usually the metric) to be present. A quantum field in a
time dependent background will generically experience particle
production - a competing effect to radiative energy loss. In fact
depending on the assumptions made, the energy loss via radiation
can be completely overwhelmed by particle production of UHECR
protons. The resulting spectrum of UHECR in such scenarios would
still have a flux at very high energies, even though each
individual proton is radiating and evolving towards low energy, as
the time dependent background pumps UHECR protons into the
spectrum at very high energies. Constraints based on the absence
of UHECR protons above a certain energy must therefore take this
particle production into account.

We have two primary goals.  The first is to illuminate in more
detail why simply deriving thresholds from an energy-momentum
conservation equation and using those thresholds to establish
constraints can be misleading. Secondly, we construct a toy field
theory model and analyze the energy loss and particle production
rates for high energy cosmic rays to estimate the magnitude of any
constraints that can realistically be established.  These
constraints turn out to be quite weak (at best).  Our focus will
be on the non-systematic analogy of the vacuum Cerenkov effect,
although some of the general conclusions will be applicable to any
reaction involving low energy particles (such as the GZK
reaction).  In more detail, section \ref{sec:premises} discusses
the problems with the rate calculation of Aloisio.  In section
\ref{sec:model} we build a model for a representative UHECR proton
(from a population at some energy) as it propagates in a
stochastic background, calculate the radiation spectrum of this
model, and explore the implications for UHECR's. We discuss
particle production and its associated complications in section
\ref{sec:heating}. Finally, in section \ref{sec:conclusions} we
summarize and speculate about possible future directions.
Throughout this paper we adopt units such that $\hbar=c=1$ and
choose metric signature $+~-~-~-$.

\section{Non-systematic dispersion and energy thresholds} \label{sec:premises}
We start by reviewing the approach taken by Aloisio to derive the
charged particle instability from the vacuum Cerenkov effect $p
\rightarrow p + \gamma$. In some preferred frame, usually taken to
coincide with the rest frame of the CMBR, the dispersion relations
assumed by Aloisio et. al. are
\begin{eqnarray} \label{eq:dispAloisio}
E^2= p^2+m^2+\eta_p \frac{p^3}{M_\textrm{Pl}}\\
E'^2= p'^2+m^2+\eta_p \frac{p'^3}{M_\textrm{Pl}}\\
\omega^2=k^2 + \eta_\gamma \frac {k^3}{M_\textrm{Pl}}
\end{eqnarray}
where $p,p',k$ are the momenta of the incoming proton, outgoing
proton, and photon respectively.  The coefficients $\eta_p,
\eta_\gamma$ are taken to be independent and randomly chosen
according to a gaussian distribution with some amplitude over
every de Broglie wavelength of the particle.  The authors motivate
this independence by noting that the time scales involved in the
interaction are all much greater than the Planck time and
therefore particles should feel different underlying quantum
gravity fluctuations. These modified dispersion relations are then
substituted into the conservation equation
\begin{equation} \label{eq:cons}
p_\mu -  k_\mu= p'_\mu.
\end{equation}
Furthermore, the authors argue that since the fluctuations are
independent the outgoing particle fluctuations can be ignored
without significantly changing the results (as for a significant
fraction of the time the outgoing particles will have Lorentz
violating coefficients much less than the incoming proton).
Squaring both sides of (\ref{eq:cons}) then leads to an equation
for the threshold momentum where the reaction is kinematically
allowed,
\begin{equation} \label{eq:insta}
p_{th} = \bigg{(} \frac {\omega m^2 M_{Pl}} {\eta_p}
\bigg{)}^{1/4}.
\end{equation}

In (\ref{eq:insta}) we have taken all the particle momenta to be
parallel as this is the appropriate configuration for a threshold
\cite{Mattingly:2002ba}. As $\omega \rightarrow 0$, the threshold
momentum also goes to zero, independent of the size of $\eta_p$,
which implies that charged particles would always be radiating low
energy photons. This is the instability as discussed in Aloisio
et. al.

We now examine the validity of this procedure.  Implicit in this
approach is the assumption that the reaction rate is fast once
kinematically allowed so that the relevant quantity for a
constraint is the threshold.  In the systematic vacuum Cerenkov
effect this is true, the energy loss rate can be shown to rapidly
approach $E^3/M_{Pl}$ for momenta only slightly above $p_{th}$.  A
GZK proton with energy $10^{19}$ eV would then radiate most of its
energy on the order of $10^{-25}$ seconds. However, the same
assumption cannot be made in the non-systematic case, especially
for low energy emitted photons. In the usual approach to
scattering in quantum field theory~\cite{Sakurai}, the
$\delta$-function constraint that enforces energy conservation is
a consequence of the \emph{adiabatic} approximation---namely that
the interaction potential (the time dependent background geometry
causing the fluctuations in this case) varies slowly over the
region of interaction. With the Aloisio et. al. assumption that
the fluctuation time scale $t_f$ is set by the energy of the
particle then $t_f\sim \frac{1}{E}$, much shorter than the
wavelength of an emitted low energy photon.  This immediately
implies that a low energy photon will interact with many
fluctuations, thereby invalidating the use of a $\delta$-function
that enforces conservation with one particular fluctuation.  More
severely, however, the adiabatic approximation fails and one will
not in general have a conserved energy. This invalidates the whole
notion of a simple threshold analysis. Instead, there will be
particle production which must be taken into account in addition
to any radiative losses.

If the random choice of the coefficients $\eta_p$ happens to add
up coherently over the interaction region, then a kinematic
threshold analysis is legitimate, as the interaction potential is
a constant over the interaction region.  The question then
becomes, is this likely over the lifetime of a UHECR with emission
of a low energy photon?  Even if the threshold formalism does not
apply in general there may be a high probability for it to apply
at least once and hence we might still have an instability. We now
estimate this probability in a simple model. First, let us
consider the simplest possible fluctuation, where $\eta_p$ takes
values $\pm 1$ on a timescale of $t_f=1/E$.  For a UHECR of energy
$10^{19}$ eV that travels over 1 Gpc, there are $10^{51}$ total
fluctuations.  If the fluctuations are to add up coherently for
some timescale given by $1/\omega$, then for $\omega=E/n$ where
$n>1$ the probability is $2^{-n}$.   Applying some simple
probability shows that for $n>O(100)$ there is almost zero chance
that the fluctuations will add up coherently at any point during
the UHECR lifetime.  This implies that a threshold approach might
only apply if we restrict the outgoing photons to have energies
above $10^{17}$ eV.  Note that even this number is conservative as
we have forced the fluctuations to have only two possible values,
thereby increasing the probability that they might be coherent.

Even in the high energy photon region of phase space we have an
issue, however. As Aloisio et. al point out, it is also assumed in
their analysis that the fluctuations in each of the particles'
energy is independent of that of the others in the spacetime
region of interaction--- that is each of the coefficients $\eta_A$
have a different independent value.  As seen however, a threshold
analysis is only applicable when dispersion fluctuations act
coherently over the entire interaction region.  If this is the
case, it is quite reasonable that the underlying spacetime
fluctuations that give rise to a coherent dispersion fluctuation
might also be coherent. The effective length scale of the
spacetime fluctuation would therefore be on the order of the
interaction region. It seems suspect to treat each dispersion
modification as independent in such as scenario, as the underlying
spacetime fluctuation is the same for all particles. In short, we
doubt that in a threshold analysis the Lorentz violating
coefficients can be considered independent.  Unfortunately,
knowledge of how they are correlated would require a more specific
model.  If the dispersion fluctuations were independent, a
threshold analysis similar to that of Aloisio might be valid,
although it would necessitate restricting the outgoing photon to
be high energy.  Such an analysis has not been performed; we shall
not do so here since we doubt an independent coefficient scenario
is applicable.

So what can one do? Note that the discussion above did not argue
that field theory fails, rather merely the implementation in terms
of a single energy-conserving $\delta$-function in the rate is not
applicable. We can still analyze the emitted radiation, we just
need to use the appropriate field theory techniques.  In the next
section we estimate the energy loss rate for low frequency
radiation ($\omega<E/100$), assuming that the UHECR dispersion
fluctuations are incoherent.  This will allow us to show that
there is in fact no stability problem for UHECR protons. We use
classical field theory for this estimation since we are
restricting to low energy outgoing photons (we can neglect any
individual photon's backreaction on the UHECR).

\section{Radiative emission in a fluctuating background}
\label{sec:model}
\subsection{General Considerations}
The vector potential from a moving particle (proton) $p$ with
charge $e$ is given in Lorentz invariant electromagnetism by the
expression
\begin{equation} \label{eq:potential}
A^\alpha(x)=\int d^4x' D_r(x-x')J^\alpha(x')
\end{equation}
where $D_r(x-x')$ is the usual retarded Green's function.
$J^\alpha(x')$ is the current, given in covariant form by
\begin{equation}
J^\alpha(x')=e\int d \tau v^\alpha(\tau) \delta^{(4)}[x'-r(\tau)]
\end{equation}
where $\tau$ is the proper time,$v^\alpha(\tau)$ is the
four-velocity and $r(\tau)$ is the proton's position.  There are
two ways that fluctuations can enter into (\ref{eq:potential}):
the Green's function or the current.  We first deal with the
Green's function.

One would naturally expect that spacetime fluctuations would
modify the Green's function for the electromagnetic potential
$A^\alpha$ (the equivalent of a the modification of the photon
dispersion in the previous section). Unfortunately, by definition
in Lorentz violating field theories there are couplings not only
to the metric, but also to the fields that are responsible for
Lorentz violation.  For example, in the aether model
of~\cite{Jacobson:2001yj}, boost invariance is broken by the
introduction of a new unit time-like vector field $u^\alpha$ which
can couple to $A^\beta$ via terms such as $u^\beta u^\gamma
F_{\alpha \beta} F_{\gamma}^\alpha$, $M_{Pl}^{-1} u^\mu F_{\mu
\alpha} (u^\beta
\partial_\beta) u^\nu \tilde{F}_\nu^\alpha $, or $M_{Pl}^{-2}
u^\beta u^\gamma F_{\alpha \beta} (u^\delta
\partial_\delta)^2 F_{\gamma}^\alpha$.
New couplings in combination with the fact that the fluctuations
yield time dependent values for the metric and $u^\alpha$ or other
Lorentz violating fields make it nigh impossible to evaluate the
electromagnetic potential with a complete Lorentz violating,
fluctuating Green's function. We can still make progress if we
assume that Lorentz violation scales with energy, as in the case
of the dispersion relations in (\ref{eq:dispAloisio}). If this is
the case, then since we are looking at low frequency emission
relative to the UHECR energy, any Lorentz violating fluctuations
in the electromagnetic sector will be much smaller than the
fluctuations in the source current. Hence, for the rest of this
discussion we will ignore Lorentz violation in the electromagnetic
sector and only consider the usual Lorentz invariant Green's
function for the electromagnetic field.

We now turn to the question of Lorentz violation in the current
$J^\alpha$. Since the current is only a function of position
(directly and through the velocity) there is actually no need to
use the fluctuating dispersion framework that has been previously
preferred. Instead, we will consider the effect of a fluctuating
velocity and relate it to dispersion relations only when comparing
with other constraints. The largest radiation response occurs for
the component of the velocity fluctuations in the direction of
propagation (which we choose to be along the z-axis), so we will
look only at this case. Finally, we must choose some model for the
fluctuations.  We shall choose random velocity fluctuations in the
z-direction with values $\pm v_f$ and some timescale $t_f$ as this
can be directly related to the fluctuating dispersion scenario.
More sophisticated choices are of course possible, for example
$v_f$ could be chosen from some underlying probability
distribution. We choose the simple bivalued random walk scenario
as it captures the essential physics without added complications.
In summary, we choose our charged particles to have a velocity in
the z-direction with magnitude $v=v_0 \pm v_f$, where the sign is
randomly chosen every $t_f$.

The deviation from the Lorentz invariant trajectory of any
particle in a population will be different, obviously, if each
particle experiences independent fluctuations. This is extremely
likely, since our observed population of UHECR's has sources that
vary over cosmological distances.  From an observational
standpoint it is irrelevant if a single UHECR proton out of a
large population experiences a significant amount of energy loss
from radiation.  The only possible observational signature is if a
large fraction of protons lose energy from fluctuations, leading
to a cutoff in the spectrum above some energy.  A population with
some initial $v_0$ (alternatively at some energy E) will spread
out over time in z around the value $z_0(t)=v_0 t$.  Since we have
chosen the random walk scenario for fluctuations, the variance of
the population is simple to calculate. Each particle deviates in
the z-direction by $z_f=\pm v_f t_f$ every step and in a time
interval $t$ there are $t/t_f$ steps, which yields a variance in z
of $v_f\sqrt{2t_f t/\pi}$.  We therefore use as a representative
trajectory
\begin{equation} \label{eq:avgtraj}
z(t)=\Big(v_0 t \pm v_f\sqrt{\frac{2}{\pi} t_f t}\Big)
\end{equation}
i.e. a particle with a deviation from the Lorentz invariant
trajectory equal to the variance of the population (assuming it
was created at $t=0$). If a particle with this trajectory radiates
much of its energy, then we know that a large fraction of the
population is radiating and vice versa.  Similarly, the velocity
corresponding to (\ref{eq:avgtraj}) is
\begin{equation} \label{eq:avgv}
v=\Big(v_0 \pm v_f \sqrt{\frac{t_f}{2 \pi t}}\Big).
\end{equation}
Equations (\ref{eq:avgtraj}) and (\ref{eq:avgv}) determine the
current $J^\alpha$.  Note that some of the particles will slow
down while others speed up since the sign in front of $v_f$ can be
either positive or negative.

As an aside, we mention that since $z(t),v$ completely determine
the representative current, our results can be quickly be
generalized to other distributions.  Models with anti-correlations
between successive fluctuations have representative trajectories
that deviate much more slowly than the random walk model we are
considering.  In general, if the rate of deviation of $z(t)$
versus time grows more slowly than $\sqrt{t}$, the corresponding
energy loss will be less (and vice versa if the deviation grows
more quickly).  Hence the constraints from energy loss on these
models are presumably weaker.  In particular the ``holographic''
model of Ng and van Dam~\cite{Ng:1993jb} would be less
constrained.  For a discussion of various models for the deviation
of particle trajectories from spacetime foam see~\cite{Ng:2004qq}.

\subsection{Energy loss rate}
Since we are considering the effect of Lorentz violation only in
the current $J^\alpha$, we can easily calculate the energy loss
rate. The energy radiated per unit frequency and solid angle
is~\cite{Jackson}
\begin{equation} \label{eq:energyloss}
\frac {d^2 I} {d \omega d \Omega} = \frac {\omega^2} {4 \pi^2}
\bigg{|}\int dt \int d^3x ~\vec{n} \times [ \vec{n} \times
\vec{J}(x,t)] e^{i \omega(t- \vec{n} \cdot \vec{x})}\bigg{|}^2
\end{equation}
where $\vec{n}$ is the unit normal between $x$ and the observation
point.  Given (\ref{eq:avgtraj}) and (\ref{eq:avgv}) the 3-current
for a particle with charge $e$ is
\begin{equation} \label{eq:current}
\vec{J}(x, t)=\hat{z} e \Big(v_0 \pm v_f \sqrt{\frac{t_f}{2 \pi
t}}\Big)\delta(x)\delta(y) \delta\Big(z-v_0 t \mp v_f
\sqrt{\frac{2}{\pi} t_f t}\Big) \Theta(t) \Theta(T-t).
\end{equation}
The Heaviside step functions $\Theta(t),\Theta(T-t)$ are present
as we are assuming the particle is created by some astrophysical
source at time $t=0$ and observed on earth at time $t=T$.
Substituting (\ref{eq:current}) into (\ref{eq:energyloss}) and
working in spherical coordinates we have
\begin{equation} \label{eq:energyloss2}
\frac {d^2 I} {d \omega d \Omega} = \frac{\omega^2 e^2} {4 \pi^2
A} sin^2 \theta \bigg{|} \int_0^T dt ~(v_0 \pm v_f
\sqrt{\frac{t_f}{2 \pi t}}) e^{i \omega t(1-v_0 \cos \theta \mp
v_f \cos \theta \sqrt{\frac {2 t_f} {\pi t}})} \bigg{|}^2
\end{equation}
where $A=1-v_0 \cos \theta$.  The integral in
(\ref{eq:energyloss2}) can be evaluated explicitly,
\begin{eqnarray}
&\int_0^T dt~ (v_0 \pm v_f \sqrt{\frac{t_f}{2 \pi t}}) e^{i \omega
t(1-v_0 \cos \theta \mp v_f \cos \theta \sqrt{\frac {2 t_f} {\pi t}})}{}\nonumber\\
&= - \frac {2e^{-\frac {i \omega B^2}{4 A}}} {\sqrt{\omega A}}
\bigg{(} \sqrt{\frac {\pi} {2}} Q \big{(} C(\sqrt{\frac {2} {\pi}}
x) + i S(\sqrt{\frac {2} {\pi}} x)
\big{)} + {}\nonumber\\
&\frac {R} {2} \big{(} \sin (x^2) - \cos (x^2) \big{)}
\bigg{)}\bigg{|}^{x=\beta}_{x=\gamma}
\end{eqnarray}
where we have defined
\begin{eqnarray}
B&=\pm v_f \cos \theta \sqrt{ \frac {2} {\pi} t_f}{}\nonumber\\
\beta&=\sqrt{\omega A T}-\sqrt{\frac{\omega B^2}{4 A}}{}\nonumber\\
\gamma&=-\sqrt{\frac{\omega B^2}{4A}}{}\nonumber\\
Q&=\frac{B}{2}(A^{-1}+(\cos \theta)^{-1}){}\nonumber\\
R&=\frac{v_0}{\sqrt{\omega A}}.
\end{eqnarray}
$C$ and $S$ are the Fresnel cosine and sine functions
respectively, defined by $C(u)+iS(u)=\int_{0}^{u} e^{\frac{i}{2}
\pi y^2}dy$.  Combining all these elements, the radiated energy is
given by
\begin{eqnarray} \label{eq:energylossfinal}
\frac {d^2 I} {d \omega d \Omega}= \frac{\omega e^2} {\pi^2 A}
\sin^2 \theta \bigg{|} \bigg{(} \sqrt{\frac {\pi} {2}} Q \big{(}
C(\sqrt{\frac {2} {\pi}} x) + i S(\sqrt{\frac {2} {\pi}} x)
\big{)} \\ \nonumber + \frac {R} {2} \big{(} \sin (x^2) - i \cos
(x^2) \big{)} \bigg{)}\bigg{|}^{x=\beta}_{x=\gamma}\bigg{|}^2
\end{eqnarray}.

\subsection{Analysis}
The radiated energy (\ref{eq:energylossfinal}) is not all due to
the effect of the fluctuations.  If we set $v_f=0$, then there is
a residual piece
\begin{equation} \label{eq:energylosstrans}
\frac {d^2 I} {d \omega d \Omega}= \frac{e^2 v_0^2} { \pi^2 A^2}
\sin^2 \theta \sin^2 \frac {\omega A T} {2}
\end{equation}
which is due to the finite time existence of the source.  We will
neglect this piece since we are looking for the energy radiated
during the particle's flight, not the energy needed to create or
destroy the source particle.  The other terms are fluctuation
dependent and reflect radiation loss over the entire travel time.

The remaining frequency and angular integrals in
(\ref{eq:energylossfinal}) are extremely complicated analytically
but numerically tractable (although difficult).  The numeric
difficulty arises because $\beta,\gamma$ can be large, which makes
the Fresnel functions and the $\sin(x^2),\cos(x^2)$ terms rapidly
oscillatory. We deal with this situation by analytically replacing
the appropriate Fresnel function with its asymptotic limit in the
relevant regions of phase space. We then rewrite the integrand
averaging over the oscillations, i.e.we replace $\sin^2 \beta^2,
\sin^2 \gamma^2$ terms
 by $1/2$ and set cross terms such as $\sin \gamma^2 \sin
\beta^2$ to zero as they are almost orthogonal.  This will
introduce small errors, however this approximation will not
significantly affect our results.

It is only possible to give the total energy loss as a function of
$\Delta_f=\pm v_f \sqrt{t_f}$, as this combination of parameters
completely controls the energy loss in (\ref{eq:energylossfinal}).
Hence we can only constrain $\Delta_f$ and not $v_f,t_f$
individually.  Since we have assumed that Lorentz violation must
scale with energy, the observationally most sensitive case is at
the high end of the UHECR spectrum, the near GZK protons, as they
have both the largest known energy and gamma factor. Furthermore,
they are presumed to be extragalactic and so travel over
cosmological distances, enhancing the effect of any small energy
loss. Hence for our energy loss estimate we take a proton of
energy $10^{19}$ eV travelling over a distance of 1 Gpc. The
energy loss as a function of $\Delta_f$ is a power law, best fit
by the curve
\begin{equation} \label{eq:EofDelta}
log~(\frac {E_{rad}} {eV})= 2.04 log~|\Delta_f \sqrt{eV}| + 51.5.
\end{equation}
There is a small deviation of a fraction of a percent in the
energy loss between the cases $\Delta_f>0$ and $\Delta_f<0$. The
difference is low because the energy loss for our parameter choice
is dominated by the $Q^2$ term in (\ref{eq:energylossfinal}) which
goes as $(\pm v_f)^2$. The important physical effect is that
protons that slow down or speed up emit in almost the same way. We
will neglect the small difference when deriving constraints.

If we temporarily assume, as has been done previously, that there
is no particle production of UHECR's from the fluctuations then we
can derive constraints on $v_f, t_f$ in the following manner. We
observe UHECR protons at energies of $10^{19}$ eV (and above). If
these protons are created at the source with an energy near
$10^{19}$ eV, then the total energy loss during their flight must
be much less than $10^{19}$ eV or else we would not see any flux
in that energy band. Demanding that $E_{rad}<10^{19}$ eV yields
the constraint $|\Delta_f|<10^{-15.9}$. It is possible, although
unlikely, that high energy protons are created with higher
energies, lose energy during flight, and reach earth at the
observed energies. If we are extremely conservative and take the
worst case scenario, that the protons that reach us at GZK
energies are created at the Planck energy, then the energy loss
can at most be the Planck energy. This yields the constraint
$|\Delta_f|<10^{-11.5}$.  We shall not consider this possibility,
instead choosing the much more plausible scenario where UHECR
cosmic rays have energies at the source near their observed
energies.

The constraint $|\Delta_f|<10^{-15.9}$ can be used to put
constraints on $v_f$, assuming some length scale for $t_f$.  There
are two length scales intrinsic to the problem, the Planck length
and the de Broglie wavelength of the UHECR proton.  If $t_f=1/E$,
as in Aloisio et. al. then $v_f<10^{-6.5}$. This constraint is
very weak in comparison with other results, as we show in section
\ref{subsec:comparison}.   If $t_f=t_{Planck}$ then the constraint
is further weakened to $v_f<10^{-1.9}$, however we remind the
reader that we have derived these constraints in a classical
framework. If the fluctuation time is less than the de Broglie
wavelength of the UHECR proton presumably the quantum nature of
the UHECR must be taken into account.  We include the $t_f =
t_{Pl}$ case simply to show that the constraint weakens as $t_f
\rightarrow t_{Pl}$. This failure does not affect our main
conclusion, that there is no way to realistically calculate useful
constraints using this method.

The constraints above only take into account frequencies up to
$E/100=10^{17}$ eV.  For informative purposes we show below the
corresponding expressions allowing $f_c=10^{18},10^{19}$ eV. These
expressions should not be considered constraints, since they might
suffer from the correlation problems discussed above. However,
they show that even if we push $f_c$ all the way to $10^{19}$ eV,
the constraints would be improved by not even two orders of
magnitude.
\begin{eqnarray}
f_c=10^{18} eV &:& log~E_{rad}= 2.07 log~|\Delta_f| + 54 \\
f_c=10^{19} eV &:& log~E_{rad}= 2.06 log~|\Delta_f| + 54.9. \\
\end{eqnarray}

\subsection{Comparison with other constraints}
\label{subsec:comparison} We now compare the strength of our
constraint, $v_f<10^{-6.5}$, with other direct constraints on
particle velocity fluctuations.  We first contrast our results
with those for systematic Lorentz violation.  The existence of
stable UHECR protons at energies of $10^{19}$ eV requires that the
speed of protons does not exceed the speed of light by more than
one part in $10^{20}$, i.e.
\begin{equation}
v_p-v_\gamma<10^{-20}
\end{equation}
which is thirteen orders of magnitude stronger than our limit on
$v_f$.  Note that our limit is, however, two-sided as both
positive and negative fluctuations emit radiation.

If we assume that the velocity fluctuations come from a modified
dispersion relation via $v=\partial E/\partial p$ then the
velocity fluctuations can be written as
\begin{equation}
v_f=\frac {(n-1) \eta_p } {2} \frac {p^{n-2}} {M_{Pl}^{n-2}}
\end{equation}
where $\eta_p$ is the O(1) coefficient for a proton.  Our
corresponding constraint on $\eta_p$ is
\begin{equation}
\eta_p<\frac {2} {n-1} \frac {10^{-6.5}} {(10^{-9})^{n-2}}.
\end{equation}
For $n>2$, as we must have since we assume Lorentz violation
scales with energy, we see that the resulting constraints are far
greater than O(1), $\eta_p< O(10^2), O(10^{11})$ for $n=3,4$
respectively.  Hence with fluctuating models there are no strong
constraints that can be placed on $n=3,4$ fluctuating dispersion,
contrary to the conclusions in~\cite{Aloisio:2002ed}.

Furthermore, our results show that there is no particle stability
crisis. Our calculation includes the contribution to the energy
loss rate from low frequency radiation and shows that it is
actually finite and small. It is true (as one can see from
(\ref{eq:energylossfinal})) that given any fluctuation in $v_f$
for any energy proton there is some radiation.  This roughly
corresponds to the statement in (\ref{eq:insta}), for any
value of $p_{in}, \eta_p$ there is some $\omega$ that is at
threshold, i.e. some radiation.  However, the loss rate is small
enough that there is no stability problem for a cosmic ray proton
over a cosmological travel time.

\section{Particle production} \label{sec:heating}
In the above analysis, we have restricted ourselves to
fluctuations in the velocity of a particle, which is enough to
calculate the radiated energy classically.  Previous work has not
taken into account the quantum nature of the UHECR protons, and in
particular the consequences of particle production. Without
particle production a continuous energy loss will actually deplete
the population of high energy protons, which is a necessary
process for the observational constraint.  Unfortunately, in a
fluctuating background this method does not work.  Systematic
Lorentz violation \footnote{We do not include more exotic
scenarios like doubly special relativity.  For an introduction to
DSR see for example ~\cite{Amelino-Camelia:2002vy}.} preserves
translation symmetry - there exists a global Killing vector by
which a conserved particle energy and number can be defined.  If
the underlying background is fluctuating, however, no such Killing
vector exists, time-translation symmetry is broken, and there will
generically be particle production.  One should expect significant
particle production in the Aloisio et. al. model since the
population of UHECR is being driven by the fluctuations on a time
scale of order of their de Broglie period.  The question is
exactly how particle production compares with the electromagnetic
loss rate. If the particle production rate is great enough, the
decrease in particle number at some energy due to radiation can be
completely compensated for. We estimate the magnitude of this
effect for the specific form of fluctuating velocity above. In
this very specific model we show that particle production is
negligible.  However, this is actually an artifact of the
idealized choice of instantaneous switching between positive and
negative fluctuations. We illustrate that in a more realistic
model the amount of particle production is wildly dependent on the
shape of the fluctuations. This makes it extremely difficult to
derive \textit{any} generic model independent constraints.

\subsection{Instantaneous random walk model}
We have assumed throughout this work a fluctuating velocity in the
z-direction and so we need a model that gives such behavior. We
will then show that tremendous amount of particle production can
occur. The simplest example is that of a massive scalar
field\footnote{Our general conclusion, that particle production
can be important, applies regardless of the bosonic or fermionic
nature of the fields. Hence we will study a massive scalar field
for simplicity.} $\phi$ minimally coupled to a background metric
of the form
\begin{equation}
g_{\alpha \beta}=\eta_{\alpha \beta} + h_{\alpha \beta}(t).
\end{equation}
$h_{\alpha \beta}(t)$ is a fluctuating term in the preferred frame
that is dependent only on time.  We assume that $h_{zz}$ is the
only non-zero component as our choice of velocity fluctuations has
been in the $z$ direction.  Furthermore we choose $h_{zz}(t)$ to
be non-zero only in some finite region $0<t<T$ so that we have
well defined asymptotic in and out states.  The equation of motion
for the scalar field in this background is
\begin{equation}
\frac{1}{\sqrt{-g}} \partial_\alpha \bigg(\sqrt{-g} g^{\alpha
\beta} \partial_\beta \phi \bigg)+m^2 \phi=0
\end{equation}
which to lowest order in $h_{zz}$ is
\begin{equation}
\Box \phi + h_{zz}(t)
\partial_z^2 \phi-\frac{1}{2}\partial_{t}h_{zz} \partial_{t} \phi
+ m^2 \phi=0.
\end{equation}
The system is translationally invariant in z, so we can assume
$\phi$ is of the form $\phi=\psi(t) e^{ipz}$. We then have the one
dimensional equation for $\psi(t)$
\begin{equation} \label{eq:scalareq2}
\partial_t^2 \psi +E_0^2  (1- h_{zz}(t))\,\psi -\partial_t h_{zz}\,\partial_{t} \psi =0
\end{equation}
where $E_0^2=p^2+m^2$.

The very specific case of the random walk scenario considered in
section \ref{sec:model} can be implemented by choosing $h_{zz}(t)$
to be
\begin{equation} \label{eq:zcomp}
h_{zz}(t)=\sum_{n=0}^{\frac{T}{t_f}-1}\Theta\,\big(t-nt_f\big)
\Theta\,\big((n+1)t_f -t\big) A_n
\end{equation}
where $A_n=\pm 2 v_f$. With this choice of $h_{zz}$, $\partial_t
h_{zz}=0$ in the middle of the fluctuations (i.e. not at the
points $t=nt_f$).  The damping term vanishes in these areas and
(\ref{eq:scalareq2}) becomes a wave equation.  To lowest order in
the small quantities $h_{zz}, m/p$ the group velocity
$v_g=\partial E/\partial p$ is
\begin{equation}
v_g=1-\frac {m^2} {2p^2} \pm v_f
\end{equation}
which shows that this model realizes the random fluctuation
scenario. The question of particle creation then becomes a
matching problem between the boundaries of the n fluctuation
regions as we transition from one wave solution to the next.

The problem can be made much simpler by changing coordinates and
eliminating the damping term entirely. We introduce a new time
variable $\eta$ defined by
\begin{equation}\label{eq:newtime}
\eta=\int \frac{dt}{1-\frac{1}{2}h_{zz}}.
\end{equation}
In $\eta$ time (\ref{eq:scalareq2}) becomes
\begin{equation} \label{eq:ho}
\ddot{\psi}(\eta)+E_0^2\Big(1-\frac{3}{4}h_{zz}^2\Big)
\psi(\eta)=0
\end{equation}
where the double-dot denotes a double time derivative with respect
to $\eta$. In this form, (\ref{eq:ho}) is simply a harmonic
oscillator with a time dependent frequency.  The key observation
is that the frequency is quadratic in $h_{zz}$.  Since we are
switching between $\pm 2v_f$ \textit{instantaneously} the field
equation in $\eta$ remains unchanged in the fluctuating region.
Hence there is only particle production at the edges of the
fluctuating region where the energy switches from $E=E_0$ to
$E_f=E_0{(1-3/4 h_{zz}^2)}^{1/2}$. Finally let us also observe
from (\ref{eq:newtime}) that for $t \leq 0\,$ $\eta=t$, while for
$t \geq T\,$ $\eta=t+\tau$ where $\tau$ is a constant.  The number
operator in $\eta$ is therefore the same as that in $t$ outside
the fluctuating region. Likewise, the asymptotic in and out vacua
are also equivalent, i.e.
\begin{equation}
a_\eta |0>_t = a_t |0>_\eta=0
\end{equation}
where $a_{\eta,t}$ are the field annihilation operators. From
these two observations it can be shown that the particle
production calculated in $\eta$ is the same as would be calculated
directly in $t$.

We have established that in this model the only points relevant
for particle production are the edges of the fluctuating region,
where $h_{zz}$ instantaneously drops to zero. The amplitude for
particle production in such a case can be calculated via the
sudden approximation \cite{Jacobson:2003vx}.  At $t=0$ (when the
particle enters the fluctuating region) the Bogoliubov
coefficients are
\begin{eqnarray}\label{eq:amp}
\alpha=\frac {1} {2}
\bigg(\sqrt{\frac{E_0}{E_f}}+\sqrt{\frac{E_f}{E_0}}\,\bigg)\\
\nonumber \beta=\frac {1} {2}
\bigg(\sqrt{\frac{E_0}{E_f}}-\sqrt{\frac{E_f}{E_0}}\,\bigg).
\end{eqnarray}
With our particular form for $E_0,E_f$ we have to lowest order in
$h_{zz}$
\begin{eqnarray}
\alpha=1 \\
\beta=\frac{3}{16} h_{zz}^2.
\end{eqnarray}

When the particle exits at $\eta=T+\tau$ this process happens
again. The Bogoliubov coefficients have the same form as
(\ref{eq:amp}) but with, but with $E_f$ and $E_0$ interchanged and
an additional phase $e^{i \phi (T+\tau)}$.  The overall rate of
particle production for an incoming N particle state can be
calculated to be
\begin{equation} \label{eq:partwalk}
N_{out}=N_{in} (1+\rho \frac {9} {256} h_{zz}^4)
\end{equation}
where $\rho$ is an O(1) coefficient that depends on the phase.
Since $|h_{zz}|=2v_f<<1$ the amplitude for particle production in
this current model is negligible.  Hence the energy loss rate from
radiation can actually be used to set a constraint on the size of
$v_f$.

\subsection{Finite time model}
The lack of significant particle production in (\ref{eq:partwalk})
can be traced back to the fact that in the random walk scenario
only the end points of the fluctuating region contribute to
particle production.  This in turn is a result of the
instantaneous fluctuations.  Realistically, of course, one might
expect some finite time $\Delta \eta$ over which $h_{zz}(\eta)$
changes. Over the course of the change there will be particle
production. To show the model dependence, we return to
(\ref{eq:ho}) which applies even if the form of $h_{zz}(\eta)$ is
not that given by (\ref{eq:zcomp}).

Consider now one fluctuation, where $h_{zz}$ changes from $2 v_f$
to $-2 v_f$ over time $\Delta \eta$.  We define the deviation from
$2 v_f$ by $2 \Delta h_{zz}=h_{zz}-2 v_f$. We further assume
$\Delta \eta<<\eta_f$ where $\eta_f$ is the fluctuation time $t_f$
in $\eta$ coordinates.  The limit $\Delta \eta \rightarrow 0$ is
therefore the random walk scenario above, where every $\eta_f$
time the fluctuations change instantaneously. If the amount of
particle production from one fluctuation is small (which we show
\textit{a posteriori}) then we can calculate it via the background
field method. We first rewrite (\ref{eq:ho}) in terms of $v_f$ and
$\Delta h_{zz}$
\begin{equation} \label{eq:ho2}
\ddot{\psi} + E'^2 \psi - 3 E_0^2\Delta h_{zz}(\Delta h_{zz} +
2v_f) \psi=0
\end{equation}
where $E'^2=E_0^2(1-3v_f^2)$.  We define $\psi=\psi_0 + \psi_1$
where $\psi_0=e^{-i E' \eta}$ satisfies the equations of motion
when $\Delta h_{zz}=0$.  We are assuming there is only a small
amount of particle production so $\psi_1<<\psi_0$. To first order
in the small quantities $\Delta h_{zz}, \psi_1$ (\ref{eq:ho2})
becomes
\begin{equation} \label{eq:ho3}
\ddot{\psi_1} + E'^2 \psi_1 = 3E_0^2 \Delta h_{zz}(\Delta h_{zz} +
2 v_f) e^{-i E' \eta}.
\end{equation}

Equation (\ref{eq:ho3}) is now a harmonic oscillator with a time
dependent force.  The particle creation from such a system is well
known (c.f. \cite{Mukhanov:2004lec}).  As an example, assume that
$\Delta h_{zz}$ is linear\footnote{This could be done by a more
complicated functional form for $h_{zz}(t)$. Our point here is not
to condone any particular model time dependence, but rather to
show that particle production cannot be ignored.} in $\eta$, i.e.
\begin{equation}
\Delta h_{zz}= \frac {-2 v_f} {\Delta \eta} (\eta - \eta_0)
\end{equation}
between times $\eta_0$ and $\eta_0 + \Delta \eta$ and zero
elsewhere. Assuming $\psi_1$ initially is zero, the outgoing
number of $\psi_1$ excitations after a single fluctuation is
\begin{equation}
N(\psi_1)=2 E_0^2 \Delta \eta^2 v_f^4
\end{equation}
and the total expectation value of the particle number is
$1+N(\psi_1)$ since $\psi_0$ was a single particle state. In the
random walk scenario, $t_f$ was set by the de Broglie wavelength
of the particle.  If we set $\eta_f$ by the same method,
$\eta_f=1/E'$, and take $\Delta \eta$ to be $\eta_f/m$ where $m$
is an integer then to lowest order $N(\psi_1)= 2 v_f^4/m$.

For an individual fluctuation the amount of particle creation is
actually small.  However, the net effect of fluctuations is
cumulative.  Significant particle creation occurs when this
cumulative effect becomes greater than or equal to the number of
initial particles (which we chose to be one in the calculation
above). After I fluctuations, the total number of particles
created can be estimated by $N_{cum} = (1+2 v_f^4/m)^I\approx 1 +
2 I v_f^4/m$. For our previous constraint of $v_f = 10^{-6.5}$ and
a proton of energy $10^{19}$ eV travelling a Gpc this works out to
be $N_{cum} \approx 10^{26} /m$.  Even if $m$ is of order $10^3$,
which implies that the fluctuations switch over a timescale 1000
times shorter than the de Broglie wavelength of the UHECR protons,
there would still be a magnification of $10^{23}$ in the number of
protons seen given an initial source flux.  Such an increase in
population is obviously unphysical, however our point is simply
that particle creation effects can completely overwhelm any
radiative losses and must be taken into account when deriving
constraints. Since the amount of particle creation is model
dependent little can be said about the expected population of
UHECR protons in a Lorentz violating fluctuating background
without a specific and concrete model for how quantum gravity
might induce such fluctuations.

\section{Conclusion} \label{sec:conclusions}
We have investigated in this paper a scenario with a fluctuating
Lorentz violating dispersion relation from an effective field
theory context.  We have found two significant new effects.  The
first is that the low frequency radiation loss from such models is
less than was calculated using threshold analyses.  The actual
constraint we get from UHECR protons is that the amplitude of the
velocity fluctuations is bounded by $10^{-6.5}$.  However, we also
have shown that almost any constraint in a fluctuating background
should be viewed with skepticism due to particle production from
the time dependent background.  For an initial one particle state,
we find that for our constraint on $v_f$, there are of order
$10^{23}$ outgoing particles, completely overwhelming any
threshold type analysis. The particle production is very model
dependent, so it seems unlikely that any generic conclusions about
fluctuating dispersion relations can be derived. It may be,
however, that the particle production itself can be used to set
limits for specific quantum gravity models.  A similar approach
has been used for causal sets in \cite{Dowker:2003hb}.

A limitation to our analysis is that we have strictly stuck to
effective field theory. One could, of course, postulate that in
some quantum gravity model such particle production does not
happen. However, then we are well outside the realm of known
physics and one must construct a new method for calculating
reaction rates, etc. that yields no particle production for
fluctuating backgrounds but also matches what we observe about
relativity and the standard model. We know of no such framework,
and hence we conclude that at the present time constraints from
models that postulate fluctuating dispersion cannot be reliably
extracted.\\

\textbf{Acknowledgments} We would like to thank Steven Carlip for
stimulating discussions. One of us (SB) would like to thank the
College of Letters and Sciences, UC Davis for a summer research
fellowship. This work was funded under DOE grant
DE-FG02-91ER40674.\\
\\

\end{document}